\documentclass[a4paper,fleqn,usenatbib]{aa}

\usepackage{txfonts}
\usepackage{pdflscape}

\usepackage{ae,aecompl}

\usepackage{graphicx}	% Including figure files
\usepackage{amsmath}	% Advanced maths commands
\usepackage{amssymb}	% Extra maths symbols
\usepackage{orcidlink}

\newcommand{\eps}{erg s$^{-1}$~}

\begin{document}

%%%%%%%%%%%%%%%%%%% TITLE PAGE %%%%%%%%%%%%%%%%%%%

% Title of the paper, and the short title which is used in the headers.
% Keep the title short and informative.
\title{ALMA observation of evolving magnetized corona in the radio-quiet changing-state AGN NGC\,1566}
\author{Arghajit Jana\inst{1}$^{\orcidlink{0000-0001-7500-5752}}$
\thanks{E-mail: arghajit.jana@mail.udp.cl},
Claudio Ricci\inst{1, 2}$^{\orcidlink{0000-0001-5231-2645}}$,
Sophie M. Venselaar\inst{1}$^{\orcidlink{0009-0001-8342-7522}}$,
Chin-Shin Chang\inst{3}$^{\orcidlink{0000-0001-9910-3234}}$,
Mai Liao\inst{1}$^{\orcidlink{0000-0002-9137-7019}}$,  \\
Yoshiyuki Inoue\inst{4,5,6}$^{\orcidlink{0000-0002-7272-1136}}$,
Taiki Kawamuro\inst{4,7}$^{\orcidlink{0000-0002-6808-2052}}$,
Franz E. Bauer\inst{8}$^{\orcidlink{0000-0002-8686-8737}}$,
Elena Shablovinskaya\inst{1}$^{\orcidlink{0000-0003-2914-2507}}$, \\
Benny Trakhtenbrot\inst{10}$^{\orcidlink{0000-0002-3683-7297}}$,
Jacob S. Elford\inst{1}$^{\orcidlink{0000-0002-6139-2226}}$,
Michael J. Koss\inst{11,9}$^{\orcidlink{0000-0002-7998-9581}}$
%
%\author{A. Jana\inst{1}
%\thanks{E-mail: arghajit.jana@mail.udp.cl}, C. Ricci\inst{1, 2}, S. M. Venselaar\inst{1}, C.-S. Chang\inst{3}, M. Liao\inst{1}, Y. Inoue\inst{4,5,6}, \\
%T. Kawamuro\inst{4,7}, F.~E. Bauer\inst{8}, E. Shablovinskaya\inst{1}, B. Trakhtenbrot\inst{10}, J. S. Elford\inst{1}, \\
%M. Koss\inst{11,9}
}
\institute{$^{1}$Instituto de Estudios Astrof\'isicos, Facultad de Ingenier\'ia y Ciencias, Universidad Diego Portales, Av. Ej\'ercito Libertador 441, Santiago, Chile \\ 
$^{2}$ Kavli Institute for Astronomy and Astrophysics, Peking University, Beijing 100871, People's Republic of China \\
$^{3}$ Joint ALMA Observatory, Avenida Alonso de Cordova 3107, Vitacura 7630355, Santiago, Chile \\
$^{4}$ Department of Earth and Space Science, Graduate School of Science, Osaka University, 1-1 Machikaneyama, Toyonaka, Osaka 560-0043, Japan \\
$^{5}$ Interdisciplinary Theoretical \& Mathematical Science Program (iTHEMS), RIKEN, 2-1 Hirosawa, 351-0198, Japan\\
$^{6}$ Kavli Institute for the Physics and Mathematics of the Universe (WPI), UTIAS, The University of Tokyo, 5-1-5 Kashiwanoha, Kashiwa, Chiba 277-8583, Japan\\
$^{7}$ RIKEN Cluster for Pioneering Research, 2-1 Hirosawa, Wako, Saitama, Saitama 351-0198, Japan \\
$^{8}$ Instituto de Alta Investigaci{\'{o}}n, Universidad de Tarapac{\'{a}}, Casilla 7D, Arica, Chile \\ 
$^{9}$ Space Science Institute, 4750 Walnut Street, Suite 205, Boulder, Colorado 80301, USA \\
$^{10}$ School of Physics and Astronomy, Tel Aviv University, Tel Aviv 69978, Israel \\
$^{11}$ Eureka Scientific, 2452 Delmer Street Suite 100, Oakland, CA 94602-3017, USA}

\date{Accepted XXX. Received YYY; in original form ZZZ}
% Enter the current year, for the copyright statements etc.
%\pubyear{2024}

% Don't change these lines
%\begin{document}

%\label{firstpage}

\abstract
{
The origin of compact millimeter (mm) continuum emission from radio-quiet AGNs (RQAGNs) is still not fully understood. Changing-state AGNs (CSAGNs) display rapid and strong variability, which can allow us to investigate the origin of the mm emission. We present here the results of the first study of the mm continuum variability of a CSAGN using archival ALMA band 6 ($\sim 230$~GHz) observations of NGC\,1566 obtained in 2014--2023. We find a positive correlation between the mm and X-ray flux with an intrinsic scatter of 0.05 dex ($1\sigma$), suggesting a common origin. The mm spectral index ($\alpha_{\rm mm}$) is found in the range of $0.13\pm0.38$ to $-0.26\pm0.53$, consistent with a compact optically thick synchrotron source. No significant correlation was found between the $\alpha_{\rm mm}$ and the mm flux. The mm/X-ray ratio also shows no clear link to the Eddington ratio but is higher in the low-accretion state. We discuss several scenarios about the origin of the mm emission in NGC\,1566. We find that synchrotron emission in the magnetized X-ray corona appears to be the most probable origin of mm emission, confirming that mm emission can be used as a tracer of AGN activity in RQAGNs.
}
%aims
%methods
%results
\keywords
{Galaxies: active --  Galaxies: nuclei -- quasars: supermassive black holes -- X-rays: galaxies --  Submillimeter: galaxies -- Accretion, accretion disks}

\titlerunning{RQ CSAGNs}
\authorrunning{Jana et al.}
\maketitle

%%%%%%%%%%%%%%%%% BODY OF PAPER %%%%%%%%%%%%%%%%%%

\section{Introduction}
\label{sec:intro}

Active galactic nuclei (AGNs), powered by accretion onto supermassive black holes (SMBHs), are classified as radio-loud (RL) or radio-quiet (RQ) based on their radio emission \citep{Begelman1984, Wilson1995}. RL AGNs exhibit strong jet-driven radio emission, while RQAGNs show weaker compact, unresolved nuclear radio sources, potentially arising from processes such as outflows, compact jets, X-ray coronae, or star formation \citep[SF; e.g.,][]{Padovani2017,Panessa2019}. Among these, the synchrotron emission from a magnetized X-ray corona as a source of radio emission is particularly intriguing \citep{Field1993,Inoue2018}, as it may also explain the heating mechanisms of the corona, which remain debated (e.g., \citealt{Beloborodov2017}, but see also \citealt{Inoue2024,Hopkins2024}).

The X-ray corona is a compact region of hot electron plasma near the SMBH, where UV seed photons are Compton up-scattered to produce X-rays \citep{Fabian2015,HM93,CT95}. Magnetic reconnection is believed to be responsible for the coronal heating \citep{DiMatteo1997,Merloni2001,Cheng2020}, and such a magnetized corona would emit synchrotron radiation detectable in the millimeter (mm) regime  \citep{DiMatteo1997,Laor2008,Inoue2014}. Theoretical models predict that the synchrotron emission would peak around $\sim 100$ GHz and extend up to $\sim 300$ GHz with a flat spectrum due to synchrotron self absorption \citep[SSA;][]{Inoue2014,Raginski2016,delPalacio2025}. If the mm emission originates in the corona, we expect it to correlate with X-ray emission strongly.

\citet{Behar2015} found a possible mm/X-ray correlation in a sample of eight RQAGNs, though this correlation weakened in a larger sample of 34 sources with a large scatter of $\sim 0.5$ dex \citep{Behar2018}.\citet{Kawamuro2022} observed a significant correlation between X-ray and 230 GHz emission while studying a sample of 98 AGNs with an intrinsic scatter of $\sim 0.34$ dex. \citet{Ricci2023mm} found a tight correlation between X-ray and 100 GHz emission with an intrinsic scatter of $\sim 0.22$ dex using high resolution ALMA data. To date, only a few studies have investigated the simultaneous mm/X-ray variability, revealing significant mm emission variability on a timescale of days \citep{Baldi2015,Behar2020,Petrucci2023,Shablovinskaya2024}. In NGC\,7469, the mm emission varied by a factor of $\sim 2$ over $\sim$ 4-5 days, while MCG\,+08-11-11 showed intra-day mm variability \citep{Petrucci2023}. Additionally,  \citet{Shablovinskaya2024} found that the mm emission in IC\,4329A changed by a factor of $\sim 3$ within four days, exceeding the variability of X-rays in the same timescale \citep[see also][for the case of GRS~1734-292]{Michiyama2024}.

Changing-look AGNs (CLAGNs) switch between type\,1 and type\,2 states, with the appearance and disappearance of broad emission lines in the UV/optical spectra on a timescale of months to years (see \citealp{Ricci2023Nat} for a recent review). These optical state transitions are mainly driven by changes in the accretion rate, and these CLAGNs are referred as changing-state AGNs \citep{Stern2018,Ricci2023Nat,Temple2023,AJ2025}. The CS transitions are often accompanied by multi-wavelength variability, including UV, optical, IR, and X-ray bands \citep{MacLeod2016,Noda2018,Zeltyn2024}.

The change of the accretion rate could occur in two possible ways: i) disk perturbations caused, for example, by a tidal disruption event \citep{Merloni2015,Trakhtenbrot2019,Ricci2020}; ii) local disk instabilities \citep{Stern2018,Noda2018}. The latter scenario seems to be the most common (e.g., \citealp{Ricci2023Nat,AJ2025}), and during these CS transitions, the geometry of the inner accretion flow is also expected to change \citep{Noda2018,Ruan2019}. In the type\,1 state, the disk extends to the inner most stable circular orbit, while, in the type\,2 state, the disk recedes, whereby the inner accretion disk is replaced by a radiatively inefficient accretion flow (RIAF) or an extended X-ray corona \citep{CT95,Reis2013,Yuan2014}.

During the CS transitions, the physical properties of the X-ray corona, such as its geometry, temperature, and electron density, are expected to evolve \citep{Noda2018,Ruan2019}. If mm emission originates from the corona, it too may vary as the AGN changes spectral state. Therefore, CSAGNs provide a unique opportunity to investigate the origin of the mm emission and its relation with the accretion state.

NGC\,1566, a nearby (z\,=0.005) CSAGN, located at a distance of $17.9$\,Mpc \citep{koss2022}, is an ideal source to investigate the origin of mm emission variability. This source has shown rapid multi-wavelength variability over decades \citep{Alloin1986,Baribaud1992}. In 2018, it underwent an outburst, transitioning to a type\,1 state with the appearance of broad optical lines, before fading back to a type\,2 state in 2020 \citep{Oknyansky2021,Ochmann2024}.

NGC\,1566 was observed in radio band ($\sim 1.4$\,GHz) in the past by ASKAP and ATCA, but at relatively in low resolution (with beam size $\sim 42''\times35''$; e.g., \citealp{Ehle1996,Elagali2019}). However, these observations are not relevant to studying the nuclear mm emission. In this work, we present the first study of mm and X-ray variability in a CSAGN, NGC\,1566, using archival high resolution ALMA Band\,6 ($\sim 230$ GHz) data spanning between 2014 and 2023. We investigate the variability and origin of the mm emission and its relation to the accretion state.

The paper is organized in the following way. In Section~\ref{sec:obs}, we present the data analysis process. In Section~\ref{sec:res}, we present our results and discuss our findings. Finally, in Section~\ref{sec:sum}, we summarize our results. Throughout the paper, we adopt $\Lambda$-CDM cosmology with $H_{\rm 0}=70$ km s$^{-1}$ Mpc $^{-1}$, $\Omega_{\rm m}=0.3$ and $\Omega_{\rm \Lambda}=0.7$. At z\,=0.005, this cosmology implies an angular scale of 0.087\,kpc/arcsec.

\begin{figure}
\centering
\includegraphics[width=8.5cm]{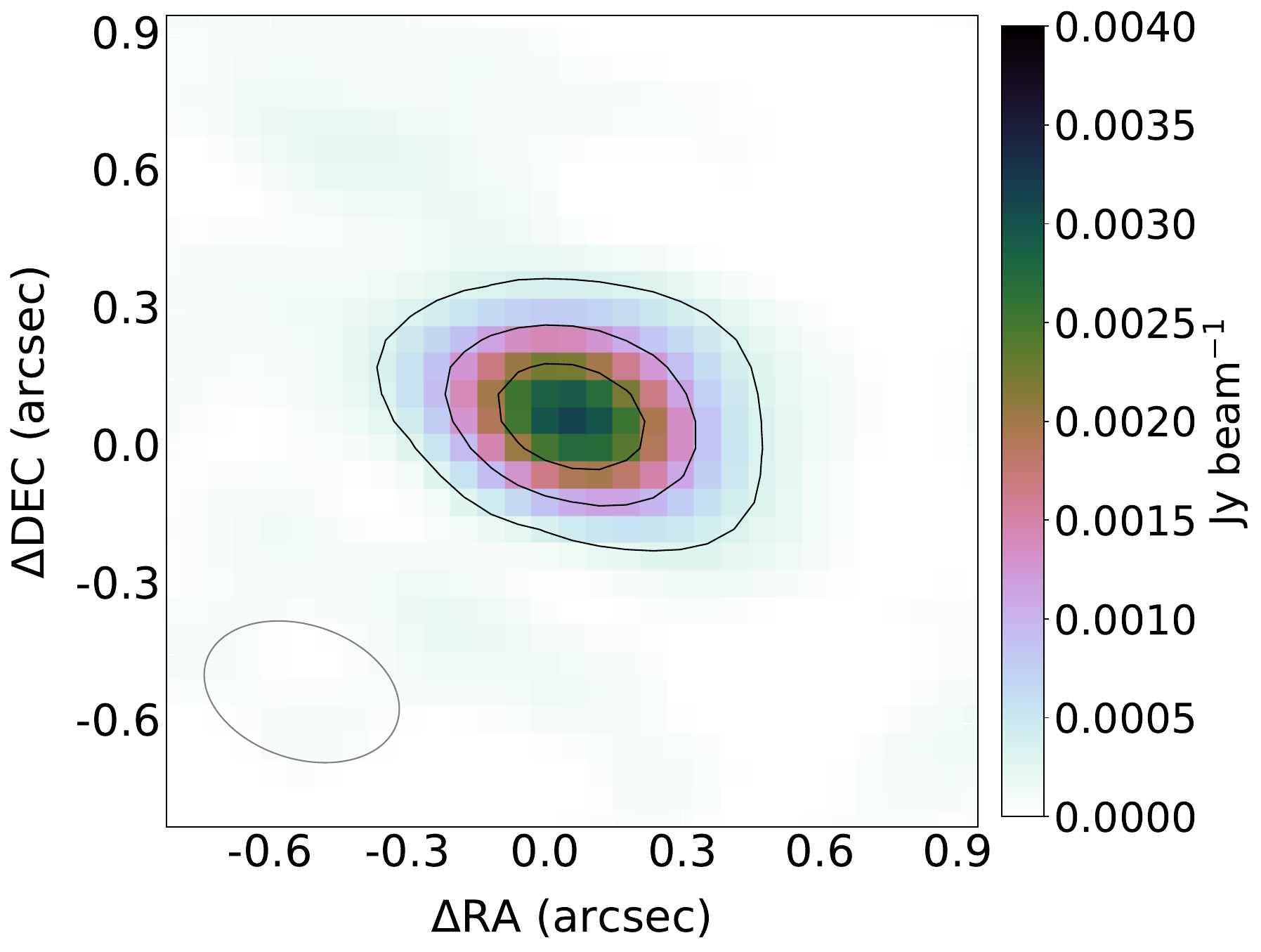}
\caption{ALMA image of NGC\,1566 in band 6 at 230 GHz, observed on 2022-06-14. The synthesized beam size (gray ellipse at the lower corner) is $0''.44 \times 0''.29$ ($38\times25$~pc). The contours of the emission are drawn at (3, 10, 20)$\sigma$, where $\sigma$\,=0.1 mJy beam$^{\rm -1}$. For NGC\,1566, 1\,arcsec angular width corresponds to 87\,pc in physical scale.}
%\caption{ALMA image of NGC\,1566 in band 6 at 230 GHz, observed on 2022-06-14. The synthesized beam size (gray ellipse at the lower corner) is $0^{`\prime\prime}.44 \times 0^{\prime\prime}.29$ ($38\times25$~pc). The contours of the emission are drawn at (3, 10, 20)$\sigma$, where $\sigma$\,=0.1 mJy beam$^{\rm -1}$. For NGC\,1566, 1\,arcsec angular width corresponds to 87\,pc in physical scale.}
\label{fig:source}
\end{figure}

\section{Observation and data analysis}
\label{sec:obs}

ALMA observed NGC\,1566 in Band 6 (230 GHz) across eight epochs between 2014 and 2023. The data were processed using CASA version 6.5.4.9 and the ALMA Pipeline version 2020.1.0.40 \citep[][please see Section~\ref{sec:alma} for details of \textsc{alma} data analysis]{Hunter2023}. For our analysis, we only considered the channels with continuum emissions for. The central source was detected at various significance levels, with a signal-to-noise ratio ranging from 100 to 700, with a position consistent with the nominal nucleus of the galaxy. Based on a Gaussian 2D fit, the central emission appears unresolved at all resolutions probed (beam sizes ranging from $0.2''$ to $7.3''$, corresponding to physical scales of $\sim 20-640$ pc). We have included an error of $10\%$ in all data, as recommended in the ALMA proposer's guide. An in-band spectral index ($\alpha_{\rm mm}$) was derived from flux measurements in four spectral windows, following $F_{\nu} \propto \nu^{\alpha_{\rm mm}}$. Details of the observations are summarized in Table\ref{tab:log}. Figure~\ref{fig:source} shows the ALMA band 6 image of the source, obtained on June 14, 2022.

The $14-150$~keV X-ray fluxes were obtained from the $14-195$ keV or $15-50$~keV {\it Swift}/BAT count rate, or $2-10$~keV {\it Swift}/XRT fluxes. For 2004–2017, the $14-195$~keV count rate was obtained from the 157-month catalog\footnote{\url{https://swift.gsfc.nasa.gov/results/bs157mon/}}. From 2018, the $15-50$~keV light curve was recovered from the BAT Hard X-ray transient monitor\footnote{\url{https://swift.gsfc.nasa.gov/results/transients/}}. The $2-10$~keV fluxes were obtained from spectral analysis of $0.5-10$~keV {\it Swift}/XRT observations. These count rates and fluxes were converted to $14-150$~keV fluxes, assuming a power-law spectrum with $\Gamma = 1.8$ \citep{Ricci17}.

The bolometric luminosity ($L_{\rm bol}$) was calculated using Eddington ratio ($\lambda_{\rm Edd}$) dependent bolometric correction factors ($\kappa_{\rm 14-195~keV}$) of \citet{Gupta2024}. According to this recent study, the bolometric correction factor is given by $\log \kappa_{\rm 14-195} = (0.13 \pm 0.04) \log \lambda_{\rm Edd} + (1.04 \pm 0.05)$.

\begin{figure*}
\centering
\includegraphics[width=17cm]{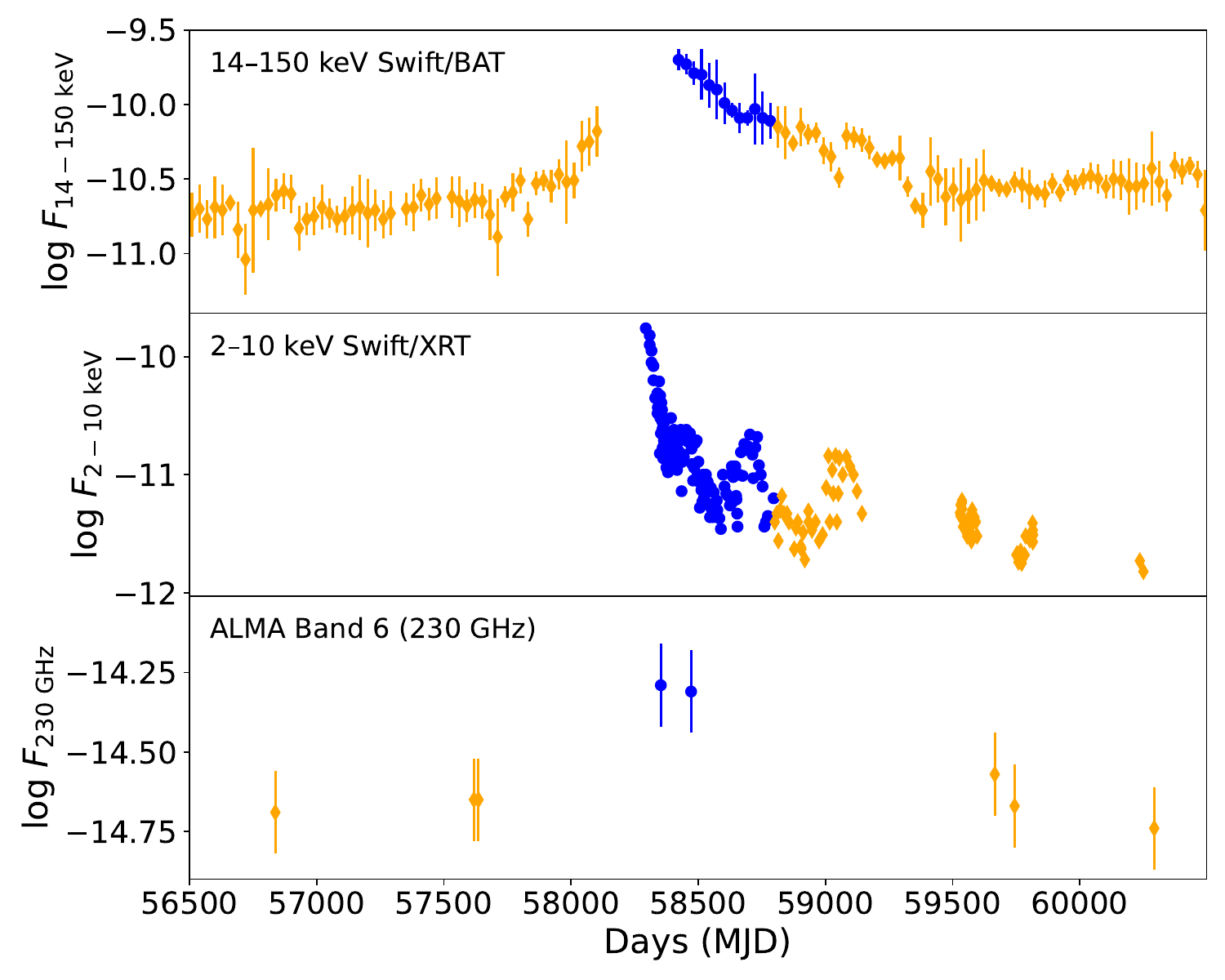}
\caption{Light curves of NGC\,1566. Top panel: {\it Swift}/BAT flux ($14-150$~keV; $F_{\rm 14-150~keV}$), middle panel: {\it Swift}/XRT flux ($F_{\rm 2-10~keV}$; $2-10$~keV), and bottom panel: ALMA band 6 flux (230 GHz; $F_{\rm 230~GHz}$). The $F_{\rm 14-150}$\,keV, $F_{\rm 2-10}$\,keV and $F_{\rm 230~GHz}$ are in unit of erg cm$^{-2}$ s$^{-1}$. The blue circles and orange diamonds represent the data from type\,1 and type\,2 states, respectively. The {\it Swift}/BAT data are obtained from the BAT 157-month catalog and BAT hard X-ray transient monitor for two periods, until 2017 and after 2018, respectively. Due to unavailability of the data, there is a gap between January 2018 and December 2018 in the {\it Swift}/BAT lightcurve. The BAT data are binned over 30 days, while the XRT data correspond to single exposures.}
\label{fig:lc1}
\end{figure*}

\section{Results and discussions}
\label{sec:res}

\subsection{Lightcurves}
\label{subsec:lc}
NGC\,1566 showed an outburst in June 2018, when the fluxes in the optical, UV, IR, and X-ray bands increased, and the source moved from type\,1.9 to type\,1 spectral states \citep{Oknyansky2019,Oknyansky2020,Parker2019,Tripathi2022}. Figure~\ref{fig:lc1} shows the light curves of NGC\,1566 in $14-150$~keV ({\it Swift}/BAT), $2-10$~keV ({\it Swift}/XRT) and 230~GHz (ALMA) in the top, middle, and bottom panels, respectively, spanning the years 2014--2023.

The $2-10$~keV X-ray and UVW2 flux increased about $\sim 25-30$ times during the outburst \citep{Oknyansky2020,AJ2021}. In optical, the `g'-band flux increased about $\sim 25\%$ compared to the pre-outburst phase \citep{Kollatschny2024}. The flux also increased in the mid-infrared (MIR) band, where the fluxes in the W1 and W2 bands increased $\sim 2.5$ and $\sim 4$ times, respectively (see Section~\ref{sec:mw_obs}). At this time, the mm flux was also observed to increase by a factor of $\sim 2$ ($\sim 9.8\pm1.0 $\,mJy). Later, in 2023, the mm flux decreased to $\sim 3.5\pm0.4$ mJy when the fluxes in other wavebands also decreased. Overall, we found that the mm continuum emission varies together with all the other bands during the CS event.

\subsection{Relation between mm and X-ray emissions}
\label{subsec:mm-Xr}
\begin{figure}
\centering
\includegraphics[width=8.5cm]{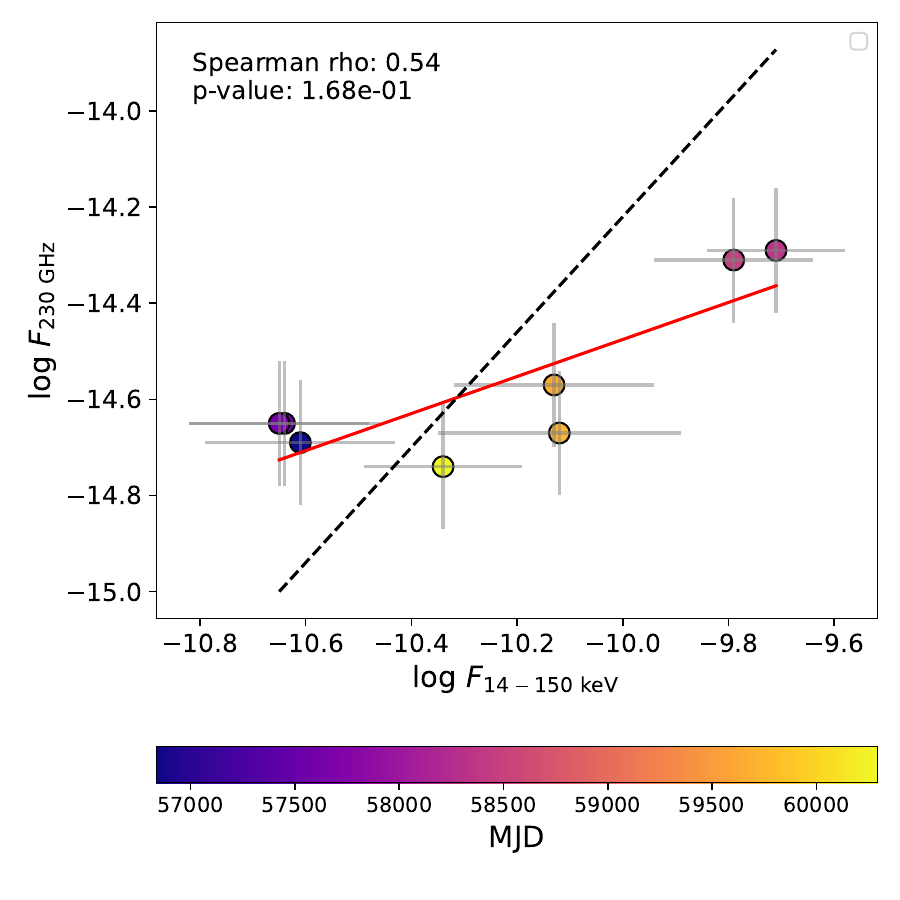}
\caption{Variation of the mm flux at 230 GHz ($F_{\rm 230~GHz}$) as a function of 14--150\,keV X-ray flux ($F_{\rm 14-150~keV}$). The red solid line represent the best-fit. The black-dashed line represent $F_{\rm 230~GHz}-F_{\rm 14-150~keV}$ relation from \citet{Kawamuro2022}.}
\label{fig:bat-mm}
\end{figure}

Figure~\ref{fig:bat-mm} displays the relation of the mm continuum flux at 230~GHz ($F_{\rm 230~GHz}$) with the X-ray continuum flux at $14-150$ keV ($F_{\rm 14-150~keV}$). We find a possible positive correlation between $F_{\rm 230~GHz}$ and $F_{\rm 14-150~keV}$ with a Spearman correlation index of 0.54 and a p-value of 0.0168. A linear fit logarithmic space yields $\log F_{\rm 230~GHz} = (0.39 \pm 0.13) \log F_{\rm 14-150~keV} + (-10.62 \pm 1.34)$, with an intrinsic scatter of 0.05 dex ($1\sigma$ level). While studying the relation between $\log L_{\rm 230~GHz}$ and $\log L_{\rm 14-150~keV}$ for 98 RQAGNs, \citet{Kawamuro2022} found a steeper slope, $\sim 1.19\pm0.08$ with a scatter of 0.30 dex. Similarly, \citet{Ricci2023mm} observed a slope of $1.22\pm 0.02$ and an intrinsic scatter of 0.22 dex for the $100$GHz mm emission ($L_{\rm 100~GHz}$) and $14-150$~keV X-ray flux in RQAGNs. Some portion of the scatter in the latter two studies could be attributed to non-simultaneous observations. The observed flatter slope in the current study could be attributed to extended or constant emission contaminating the weaker flux measurement.

\begin{figure}
\centering
\includegraphics[width=8.5cm]{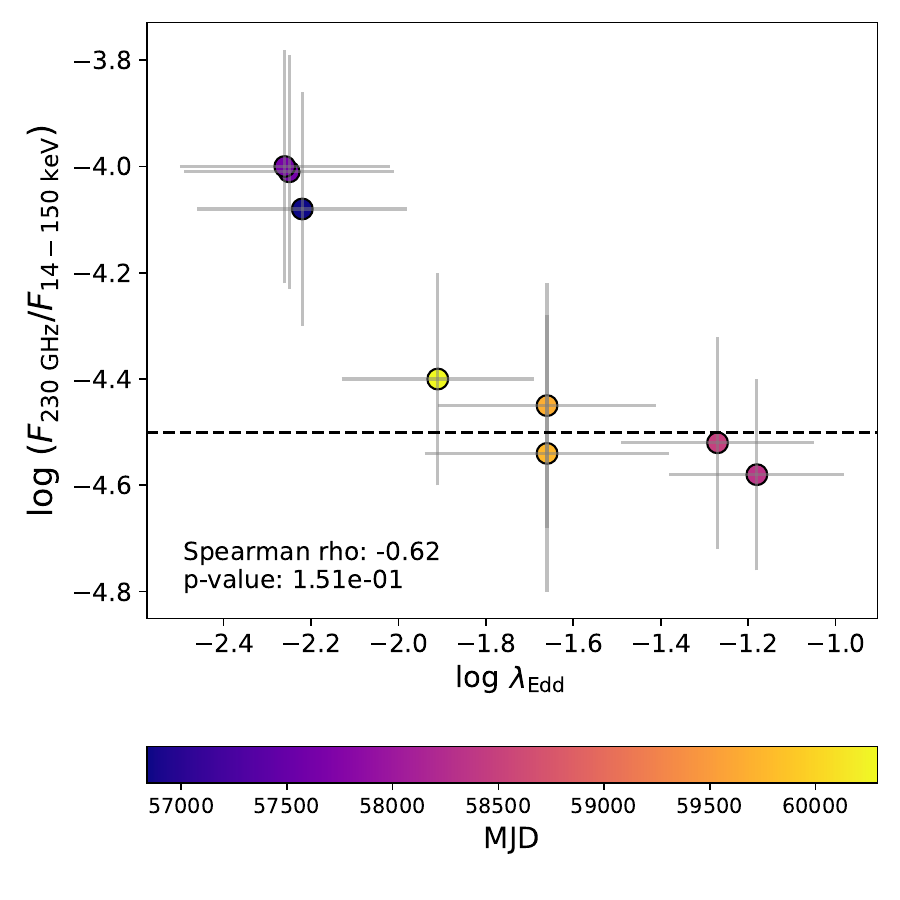}
\caption{Variation of the ratio of the mm flux to X-ray flux as a function of Eddington ratio. The horizontal dashed line represents the ratio of the mm flux to X-ray flux from \citet{Kawamuro2022}.}
\label{fig:edd_ratio}
\end{figure}

Figure~\ref{fig:edd_ratio} shows the relation between $F_{\rm 230~GHz}/F_{\rm 14-150~keV}$ and $\lambda_{\rm Edd}$. The median of the ratio between mm and X-ray flux is found to be $\log (F_{\rm 230~GHz}/F_{\rm 14-150~keV}) = -4.42 \pm 0.21$. This result is consistent with the previous study of \citet{Kawamuro2022}. The ratio is found to be somehow higher at low $\lambda_{\rm Edd}$: for $\lambda_{\rm Edd} < 10^{-2}$, the median ratio is found to be $-4.01\pm 0.22$, while for $\lambda_{\rm Edd} > 10^{-2}$ it is $-4.52\pm 0.20$.

We did not observe any correlation between $\alpha_{\rm mm}$ and the $F_{\rm 230~GHz}$ (Figure~\ref{fig:alpha}). NGC\,1566 generally showed a flat slope ranging between $-0.3\pm0.5$ and $0.1\pm0.4$, which indicates a compact optically thick source of mm emission \citep{Rybicki1979}.

%%%%%%%%%%%%%%%

\subsection{Origin of the mm emission}
\label{subsec:origin}

\subsubsection{Star formation and dust}
\label{subsec:sf-dust}
Nuclear star formation (SF) can contribute to the mm emission through mechanisms such as free-free emission from \ion{H\,II} regions, synchrotron radiation from cosmic rays, or thermal dust emission \citep{Condon1992,Tabatabaei2017}. The nuclear ($<1''$) SF rate of NGC\,1566 is estimated to be $5.3\times 10^{-3}~{\rm M}_{\odot}$ yr$^{-1}$ \citep{Smajic2015}. We estimated the contribution of SF to the mm emission at 230~GHz due to the free-free and synchrotron emission using Equation\,3 of \cite{Kawamuro2022}. We obtained $L_{\rm 230~GHz}^{\rm SFR} \approx 5.6 \times 10^{33}$ \eps ($\approx 0.07$ $\mu$Jy), which is four orders of magnitude lower than the observed mm emission.

Thermal dust emission typically produces a steep spectral index \citep[$\alpha_{\rm mm} \sim 3.5$;][]{Mullaney2011}, which is inconsistent with the flat slope ($-0.26$ to $0.13$) observed in NGC\,1566. Furthermore, dust emission dominates at higher frequencies ($\nu > 300$ GHz), making it unlikely to contribute significantly at 230 GHz \citep{Barvainis1992,Hughes1993,Draine2007,Baskin2021}. At $\sim 230$ GHz, expected free-free emission would be $\sim 1$~mJy, which is a few times less than the observed flux \citep{Baskin2021}. Additionally, the free-free emission from a dusty torus and SF would not exhibit variability on the observed timescales \citep{Baskin2021}

\subsubsection{Compact jet}
\label{subsec:jet}

A sub-parsec-scale unresolved jet could, in principle, contribute to the mm emission observed in NGC\,1566 \citep[e.g.,][]{Panessa2019}. Compact jets are well studied in black hole X-ray binaries (BHXBs), where their presence and evolution are strongly tied to the accretion state of the system \citep{Corbel2000,Fender2004,Russell2015,Tetarenko2021}. In BHXBs, a steady, compact jet produces strong radio emission during the low-hard state, whereas in the high-soft state, the jet weakens, leading to a significant drop in radio flux \citep{Fender2004,AJ2017,Tremou2020}.

NGC\,1566 is a CSAGN, which showed dramatic spectral transitions. The CSAGNs often show striking similarities in the spectral and timing properties with the BHXBs \citep{Noda2018,Ruan2019,AJ2021}. If a similar disk-jet coupling remains the same in CSAGNs, one would expect a decline in jet-related mm emission during high-accretion states, mirroring the behavior seen in BHXBs. However, in NGC\,1566, the mm flux increases with the accretion activity, contradicting this expectation. This indicates that compact jets are unlikely to be the primary contributor to the mm emission of NGC\,1566.

\subsubsection{Outflow driven shock}
\label{subsec:outflow}

The AGN outflows can collide with the surrounding interstellar medium and produce shocks. The shock accelerates electrons, which in turn produce synchrotron emission (e.g., \citealp{Hwang2018,Liao2024}). It has been suggested that the ratio of synchrotron to AGN bolometric flux would be $\sim 10^{-5}-10^{-6}$ if $\sim 0.5\%-5\%$ of the bolometric luminosity is converted into the kinetic energy, and then $\sim 1\%$ of the outflow energy is used to produce relativistic particles that radiate synchrotron emission \citep{Nims2015}. In this case, the radio emission is expected to arise from a size of several 100\,pc \citep{Yamada2024}.

Interestingly, we found a similar ratio between the mm and bolometric flux ($F_{\rm 230~GHz} \sim 10^{-5}F_{\rm bol}$) in NGC\,1566. NGC\,1566 showed X-ray wind with velocity, $v_{\rm out} \sim 500$ km s$^{-1}$ in the high accretion state \citep{Parker2019}. Conservatively assuming the outflow velocity as the escape velocity, we estimated the location of the outflow as $> 7\times 10^{14}$ cm $\sim 2\times 10^{-4}$ pc. The mass outflow rate can be estimated using $\dot{M}_{\rm out}=\Omega m_{\rm p} n_e v_{\rm out} R^2$ \citep{Krongold2007}, where $\Omega$ is the covering factor of the wind, $n_e$ is the electron density. 
Substituting, $N_{\rm H}=nR$ in the above equation, we obtained, $\dot{M}_{\rm out}=\Omega m_{\rm p} n_e v_{\rm out} R^2$, where $N_{\rm H}\sim 2.5\times10^{20}$\,cm$^{-2}$ \citep{Parker2019}. We estimated a conservative value of mass outflow rate as $\dot{M}_{\rm out} \sim 2 \times 10^{19}$ g s$^{-1}$. This would corresponds to a kinetic energy of $E_{\rm kin}=1/2 \dot{M}_{\rm out}v_{\rm out}^2 \sim 10^{34}$ \eps which translates to the flux of $\sim 2$\,$\mu$Jy at 230\,GHz. This is lower by several orders than the observed continuum mm emission in NGC\,1566.

\subsubsection{Synchrotron emission in X-ray corona}
\label{subsec:corona}

A magnetized X-ray corona, heated by magnetic reconnection, can naturally produce mm emission via synchrotron radiation from relativistic non-thermal electrons \citep{DiMatteo1997, Fabian2002,Laor2008,Inoue2014,Raginski2016}. The observed tight correlation between the mm and X-ray flux suggests that mm and X-ray emission could be coupled and originate in a similar region \citep{Kawamuro2022,Ricci2023mm}.

We find no correlation between $\alpha_{\rm mm}$ and $F_{\rm 230~GHz}$, which is consistent with and in agreement with the findings of \citet{Kawamuro2022}. 
The observed $\alpha_{\rm mm}$ suggests that the mm emission originates in an optically thick region. We calculated the size of the self-absorbed synchrotron source as $R_{\rm mm}\sim 4-7 \times 10^{-5}$ pc or $\sim 100-200$~$R_{\rm g}$, considering magnetic energy density is in equipartition with the photon energy density, and using Equation\,(22) of \citet{Laor2008}. The equipartition magnetic field ($B_{\rm eq}$) is found in the range $\sim 100-300$\,G, which is higher than typical value of magnetic field \citep[e.g.,][]{Inoue2018}. Considering magnetic field B=10\,G, as seen in IC\,4329A (\citealp{Inoue2018}, see also \citealp{Inoue2020,Michiyama2023}), the size of the mm emitting source would be $\sim 50-100$ $R_{\rm g}$. The size of the X-ray corona ($R_{\rm X}$) is estimated to be $R_{\rm X}\sim 20-40$ $R_{\rm g}$ from the broadband X-ray spectral study of NGC\,1566 \citep{AJ2021}, using \textsc{optxagnf} model \citep{Done2012}. This indicates the radio sphere may coincide with the X-ray corona or be slightly extended. Considering $R_{\rm X}=R_{\rm mm}$, we estimated the magnetic field in the range of $\sim0.1-1$\,G. This is similar to the magnetic field found in lensed quasar RXJ\,J1131--1231, where the mm emitting region is constrained to be $<46~R_{\rm g}$ \citep{Rybak2025}. Recently, \cite{delPalacio2025} constructed radio to far infrared SED model for coronal mm emission to constrain the magnetic field and coronal size. By applying their model to seven RQAGNs, the authors found B in the range of $\sim 10-150$\,G, which is higher than our simple estimation. This suggests a detailed SED modeling is required to constrain the magnetic field in NGC\,1566.

\begin{figure}
\centering
\includegraphics[width=8.5cm]{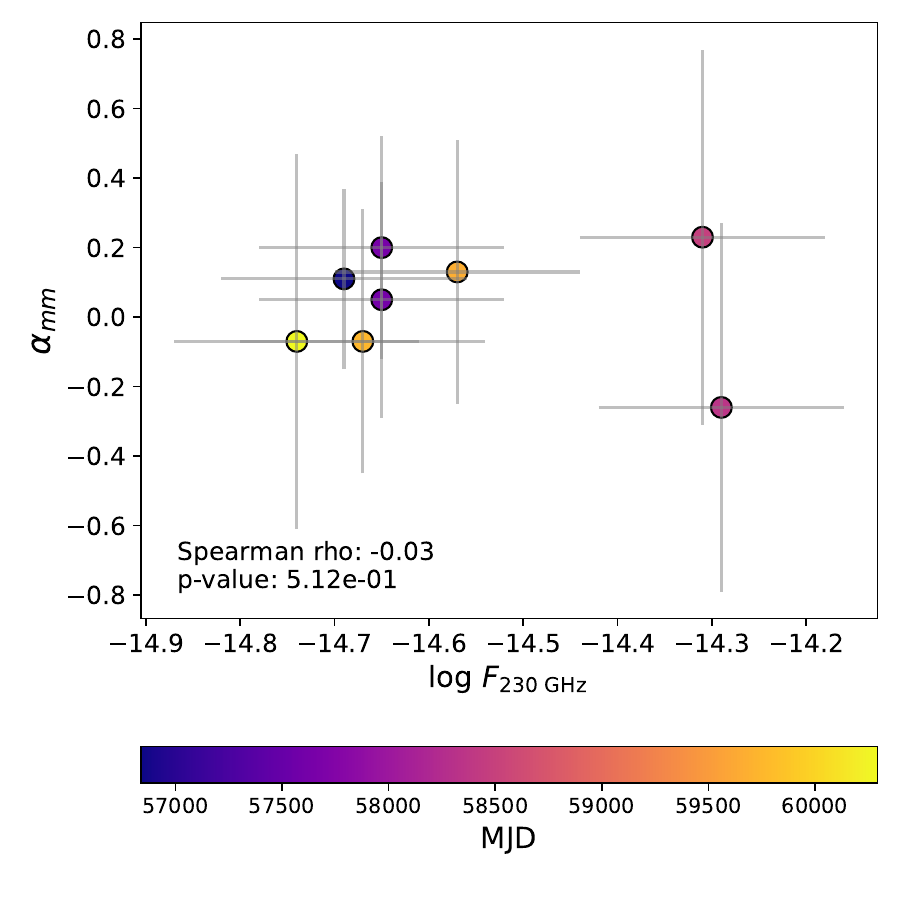}
\caption{Variation of in-band spectral index ($\alpha_{\rm mm}$) as a function of 230~GHz flux ($F_{\rm 230~GHz}$). The red solid line shows the best-fit.}
\label{fig:alpha}
\end{figure}

The enhancement of mm flux compared to the X-ray flux at low Eddington ratios (see Fig~\ref{fig:edd_ratio}) can be attributed to two mechanisms: an increase in the non-thermal electron population or strengthening of the magnetic field \citep{Inoue2014}. The observed constant spectral index ($\alpha_{\rm mm}$, Fig.~\ref{fig:alpha}) indicates minimal change in the electron energy distribution, suggesting cooling effects are not significant. This stability in $\alpha_{\rm mm}$ makes substantial magnetic field variations unlikely. The observed possible anti-correlation of the mm to X-ray emission ratio with $\lambda_{\rm Edd}$ (Fig.~\ref{fig:edd_ratio}) provides an interesting insight. Lower $\lambda_{\rm Edd}$, which implies lower accretion densities, correspond to higher mm fluxes. This can be explained by the density dependence of particle acceleration efficiency -- in lower-density flows, particles are more easily accelerated to non-thermal energies. Consequently, this could lead to a larger non-thermal electron population and enhanced mm emission. Further dense light curve data sets would test this scenario.

\subsection{Implication of mm emission in NGC\,1566}
\label{subsec:evo}

NGC\,1566 underwent an outburst in 2018, transitioning to the type\,1 state from the type\,2 state \citep{Oknyansky2019,Ochmann2024}. Later in 2020, the source moved back to the type\,2 state with the fluxes decreasing in all wavebands \citep{Xu2024}. The change in the accretion rate drives the CS transition in this source \citep{AJ2021,AJ2025}. The X-ray flux, which is considered to be a tracer of the accretion rate in AGN, is found to correlate with the UV, optical, and MIR emission \citep{Oknyansky2019,Oknyansky2020}. The mm emission is also found to follow a similar trend as the UV, optical, MIR, and X-ray emission. The tight correlation between X-ray and mm emission in NGC\,1566 suggests a strong link between mm emission and accretion activity.

A simple estimation indicates that the free-free emission from dust could contribute up to $\sim 1$ mJy at $\sim 230$\,GHz in NGC\,1566, making this contribution non-negligible (see Section~\ref{subsec:sf-dust}). Despite this potential contamination, various correlations between mm and X-ray emission indicate that the X-ray corona remains the dominant source of mm emission in NGC\,1566. Our results confirm the idea that the mm emission can be used as a tracer of AGN activity in RQAGNs \citep[see also,][]{Kawamuro2022,Ricci2023mm}.

\section{Summary \& Conclusions}
\label{sec:sum}
In this work, we investigate the mm emission variability of CSAGN for the first time, providing new insights into its nature. Here, we studied NGC\,1566 using archival high-resolution ($0''.3-7''.3$) ALMA observations in Band\,6 ($\sim 230$~GHz) taken in eight different epochs between 2014 and 2023. Our findings are the following:

\begin{enumerate}
\item We observed a strong correlation between mm and X-ray emission with an intrinsic scatter of 0.05 dex ($1\sigma$), suggesting they are energetically coupled and likely originate in a similar physical region.

\item We discuss several possible physical mechanisms that could be responsible for the bulk of the mm continuum emission in NGC\,1566. We discard star formation, outflow-driven shock, and compact jet as possible contributors to the mm emission. The free-free emission from dust could contribute up to $\sim 1$ mJy in mm wavebands; however, this would be non-variable. Despite this, the X-ray corona is found to be a dominant contributor to the nuclear mm emission.

\item We did not observe any significant correlation between $\alpha_{\rm mm}$ and $F_{\rm 230~GHz}$, although a possible negative trend is seen. The observed $\alpha_{\rm mm}$ suggests that the mm emission arises from a compact optically thick radio sphere, which could coincide with the X-ray corona.

\item The median of $\log (F_{\rm 230~GHz}/F_{\rm 14-150~keV})$ is found to be $-4.42 \pm 0.21$. While no strong correlation was found between the flux ratio and $\lambda_{\rm Edd}$, the ratio appears higher in the low-accretion state. This could be due to more efficient particle acceleration in low-density flows. This results in a larger non-thermal electron population and enhanced mm emission compared to the X-ray emission.

\item The mm emission follows a similar trend as the X-ray, UV, optical, and MIR wavebands. Along with the strong mm-X-ray correlation, this suggests a close connection between mm emission and the accretion rate. These findings support the use of mm emission as a tracer of accretion activity in radio-quiet AGNs.

\end{enumerate}

Our findings indicate that the compact X-ray corona is the most probable source of mm-wave emission in NGC\,1566. Future studies will focus on observing the source across different accretion states using simultaneous multi-wavelength data, including radio, mm, optical, and X-ray bands. 
Long-term monitoring campaigns in the mm-wave band will be crucial for capturing transient phenomena and understanding the connection between mm-wave emission and changes in the accretion flow.
It will also help us to understand the radio emission in RQAGNs, as well as the heating mechanism of the X-ray corona.

\begin{acknowledgements}
We thank the reviewer E. Behar for his invaluable comments and suggestions that improved the paper significantly.
We acknowledge support from ANID grants 
FONDECYT Postdoctoral fellowship 3230303 (AJ), 
Fondecyt Regular grant 1230345 (CR) and 1241005 (FEB), 
ANID BASAL project FB210003 (CR, FEB), the China-Chile joint research fund (CR), and the Millennium Science Initiative, AIM23-0001 (FEB).
YI is supported by NAOJ ALMA Scientific Research Grant Number 2021-17A; JSPS KAKENHI Grant Number JP22K18277; and World Premier International Research Center Initiative (WPI), MEXT, Japan.
TK is supported by JSPS KAKENHI 23K13153.
BT is supported by the European Research Council (ERC) under the European Union's Horizon 2020 research and innovation program (grant agreement number 950533).
This paper makes use of the following ALMA data: ADS/JAO.ALMA\#2012.1.00474.S, ADS/JAO.ALMA\#2015.1.00925.S, ADS/JAO.ALMA\#2017.1.00392.S, ADS/JAO.ALMA\#2018.1.01651.5, ADS/JAO.ALMA\#2021.1.01150.S, ADS/JAO.ALMA\#2023.1.01182.S.
ALMA is a partnership of ESO (representing its member states), NSF (USA) and NINS (Japan), together with NRC (Canada), NSTC and ASIAA (Taiwan), and KASI (Republic of Korea), in cooperation with the Republic of Chile. The Joint ALMA Observatory is operated by ESO, AUI/NRAO and NAOJ.
\end{acknowledgements}
%\section*{Acknowledgements}

\section*{DATA AVAILABILITY}
All the data used in the paper are publicly available.

\bibliographystyle{aa}
\bibliography{ngc1566_alma}

%%%%%%%%%%%%%%%%%%%%%%%%%%%%%%%%%%%%%%%%%%%%%%%%%%

%%%%%%%%%%%%%%%%% APPENDICES %%%%%%%%%%%%%%%%%%%%%
\appendix

\section{Observation log}
\label{sec:log}

\begin{table*}[!ht]
\centering
\caption{Details of the ALMA Observations}
\begin{tabular}{llllllllllll}
\hline
\hline
Epoch & Project code & $\nu_{\rm cen}$ & Date & Date & $\theta^{\rm maj}_{\rm beam} \times \theta^{\rm min}_{\rm beam}$ & $\theta^{\rm maj}_{\rm beam,pc} \times \theta^{\rm min}_{\rm beam. pc}$ &  $S_{\rm \nu}^{\rm peak}$ &  $S_{\rm \nu, 7.31'' \times 4.96''}^{\rm peak}$ & SNR \\
      &    & (GHz)  & (yyyy-dd-mm) & (MJD) & (arcsec) & (pc) &  \\
(1) & (2) & (3) & (4) & (5) & (6) & (7) & (8) & (9) & (10) \\ 
\hline
1	&	2012.1.00474.S	&	235.37	&	2014-06-29	&	56837	&	$0.63	\times	0.55$	&	$55	\times	48$ &	3.90	&	3.69	& 177\\
2	&	2015.1.00925.S	&	223.81	&	2016-08-18	&	57618	&	$7.31	\times	4.85$	&	$636\times	422$&	3.91	&	4.17	& 333\\
3	&	2015.1.00925.S	&	223.79	&	2016-09-05	&	57636	&	$7.08	\times	4.96$	&	$616 \times	432$&	4.22	&	4.32	& 313\\
4	&	2017.1.00392.S	&	223.63	&	2018-08-23	&	58353	&	$1.04	\times	0.77$	&	$90	\times	67$	&	9.77	&	10.88	& 579\\
5	&	2018.1.01651.5	&	223.66	&	2018-12-20	&	58472	&	$0.96	\times	0.79$	&	$84	\times	69$	&	9.30	&	10.54	& 705\\
6	&	2021.1.01150.S	&	234.82	&	2022-03-28	&	59666	&	$1.24	\times	1.07$	&	$108\times	93$	&	5.07	&	5.20	& 256\\
7	&	2021.1.01150.S	&	234.82	&	2022-06-14	&	59744	&	$0.44	\times	0.29$	&	$38	\times	25$	&	4.05	&	4.36	& 226\\
8	&	2023.1.01182.S	&	224.93	&	2023-12-15	&	60293	&	$0.27	\times	0.23$	&	$23	\times	20$	&	3.48	&	3.65	& 171\\
\hline
\hline
\end{tabular}
\leftline{Note: (1) Epoch of observation, (2) ALMA project code, (3) central frequency in GHz, (4) and (5) observation date in}
\leftline{yyyy-mm-dd and MJD unit respectively. (6) Beam size in unit of arcseconds along major and minor axes, (7) physical}
\leftline{beam size in unit of parsecs along major and minor axes, (8) peak flux density with original beam size in mJy beam$^{-1}$,}
\leftline{(9) peak flux density with $7.31'' \times 4.96''$ beam size in mJy beam$^{-1}$, (10) signal-to-noise
ratio of the detection.}
\leftline{We used $10\%$ uncertainties for flux density in all epochs.}
\label{tab:log}
\end{table*}

\begin{table}[!ht]
\centering
\caption{Important physical parameters}
\begin{tabular}{ccccccccc}
\hline
\hline
%Epoch & $\log F_{\rm 230~GHz}$ &  $\log F_{\rm 14-150~keV}$ &  $\log (F_{\rm 230~GHz}/F_{\rm 14-150~keV})$ & $\log \lambda_{\rm Edd}$ & $\alpha_{\rm mm}$ & $R_{\rm mm}$ & $R_{\rm mm}$ & $B_{\rm eq}$\\ 
Epoch & $\log F_{\rm 230~GHz}$ &  $\log F_{\rm 14-150~keV}$ &  $\log (\frac{F_{\rm 230~GHz}}{F_{\rm 14-150~keV}})$ & $\log \lambda_{\rm Edd}$ & $\alpha_{\rm mm}$ & $R_{\rm mm}$ & $R_{\rm mm}$ & $B_{\rm eq}$\\  
    &  (erg cm$^{-2}$ s$^{-1}$) & (erg cm$^{-2}$ s$^{-1}$) &   &   &   & ($10^{-5}$~pc) & ($R_{\rm g}$) & (G) \\
(1) & (2) &   (3) & (4) & (5) & (6) & (7) & (8) & (9)  \\ 
\hline
1	&$	-14.69	\pm	0.13	$&$	-10.61	\pm	0.18	$&$	-4.08	\pm	0.22	$&$	-2.22	\pm	0.24	$&$	0.11	\pm	0.26	$&$	3.84	$&$	117	$&$	143	$	\\
2	&$	-14.65	\pm	0.13	$&$	-10.64	\pm	0.18	$&$	-4.01	\pm	0.22	$&$	-2.25	\pm	0.24	$&$	0.05	\pm	0.34	$&$	3.93	$&$	120	$&$	133	$	\\
3	&$	-14.65	\pm	0.13	$&$	-10.65	\pm	0.17	$&$	-4.00	\pm	0.22	$&$	-2.26	\pm	0.24	$&$	0.20	\pm	0.32	$&$	3.94	$&$	120	$&$	132	$	\\
4	&$	-14.29	\pm	0.13	$&$	-9.71	\pm	0.13	$&$	-4.58	\pm	0.18	$&$	-1.18	\pm	0.20	$&$	-0.26	\pm	0.53	$&$	7.03	$&$	214	$&$	257	$	\\
5	&$	-14.31	\pm	0.13	$&$	-9.79	\pm	0.15	$&$	-4.52	\pm	0.20	$&$	-1.27	\pm	0.22	$&$	0.23	\pm	0.54	$&$	6.78	$&$	206	$&$	240	$	\\
6	&$	-14.57	\pm	0.13	$&$	-10.13	\pm	0.19	$&$	-4.45	\pm	0.23	$&$	-1.66	\pm	0.25	$&$	0.13	\pm	0.38	$&$	4.84	$&$	147	$&$	213	$	\\
7	&$	-14.67	\pm	0.13	$&$	-10.12	\pm	0.23	$&$	-4.54	\pm	0.26	$&$	-1.66	\pm	0.28	$&$	-0.07	\pm	0.38	$&$	4.44	$&$	135	$&$	234	$	\\
8	&$	-14.74	\pm	0.13	$&$	-10.34	\pm	0.15	$&$	-4.40	\pm	0.20	$&$	-1.91	\pm	0.22	$&$	-0.07	\pm	0.54	$&$	3.94	$&$	120	$&$	198	$	\\
\hline
\hline
\end{tabular}
\leftline{Note: (1) epochs, (2) and (3) $230$ GHz and $14-150$ keV flux in unit of erg cm$^{-2}$ s$^{-1}$, respectively. (4) Ratio of mm flux}
\leftline{to X-ray flux, (5) Eddington ratio, (6) in-band spectral index, (7) and (8) size of mm emitting region in 10$^{-5}$ pc and}
\leftline{ $R_{\rm g}$, respectively. (9) Equipartition magnetic field in Gauss.}
\label{tab:data}
\end{table}

\section{Multi-wavelength observations}
\label{sec:mw_obs}
\begin{figure}
\centering
\includegraphics[width=8.5cm]{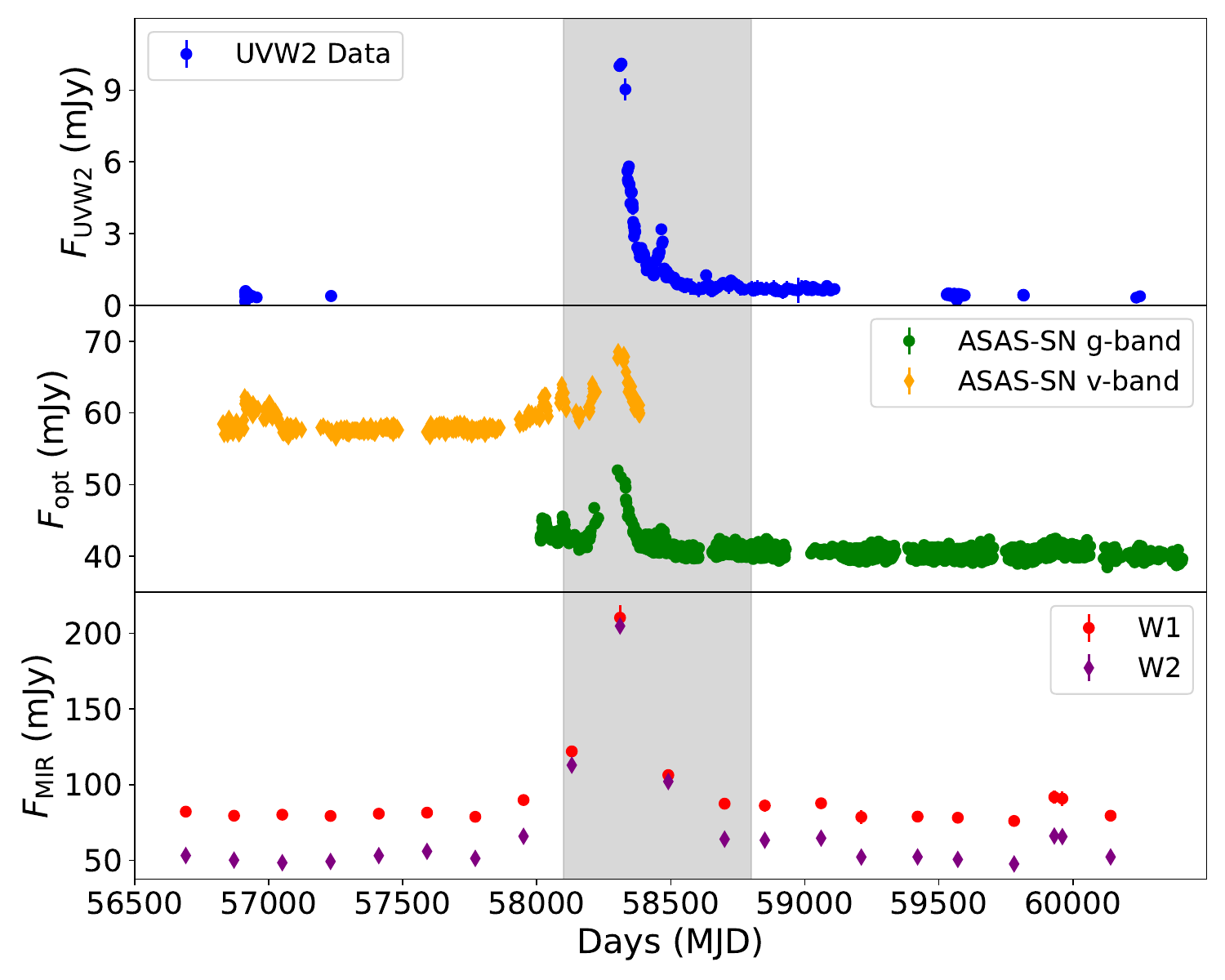}
\caption{Lightcurves of NGC\,1566. Top panel: UV flux from {\it Swift}/UVOT in UVW2 band ($F_{\rm UVW2}$). Middle panel: optical flux ($F_{\rm opt}$) from ASSAS-Sn
%\footnote{\url{https://asas-sn.osu.edu/photometry}}.}
The orange diamonds and green circles represent the lightcurve in `v' and `g' band, respectively. Bottom panel: Mid-infrared fluxes ($F_{\rm MIR}$), obtained from WISE and NEOWISE. The red circles and purple diamonds represent the lightcurve in W1 and W2 bands, respectively.}
\label{fig:lc2}
\end{figure}

The {\it Swift}/UVW2 photometric flux was estimated using \textsc{uvotsource} task. The AGN flux was estimated by subtracting the host galaxy contribution in the UVW2 band. The host galaxy flux was taken from \citet{Gupta2024}. We obtained mid-infrared (MIR) lightcurves from WISE in W1 (3.4 $\mu m$) and W2 (4.6 $\mu m$) band from IPAC. We also obtained optical lightcurves in \textsc{g} and \textsc{v} band from ASSAS-Sn archive\footnote{\url{https://asas-sn.osu.edu/photometry}}.
. Figure~\ref{fig:lc2} shows the lightcurve in UVW2, optical (v and g-band) and MIR wavebands.

\section{ALMA data reduction}
\label{sec:alma}
We utilized archival ALMA band-6 observations, which was obtained in eight epochs. 
The data were calibrated using python script \texttt{scriptForPI.py} in appropriate CASA version.
Once the data were calibrated, and measurement set (MS) was generated, we used CASA 6.5.4.9 for the following tasks.
First, using CASA task \texttt{tclean}, we generated clean images with appropriate cell sizes with weighting = briggs (robust=0.5) and original beam sizes. We used all the spectral windows to generate each clean image.
With the same process, the clean images for individual spectral windows are also generated, which we used to calculate the spectral index ($\alpha_{\rm mm}$).

\begin{figure}
\centering
\includegraphics[width=8.5cm]{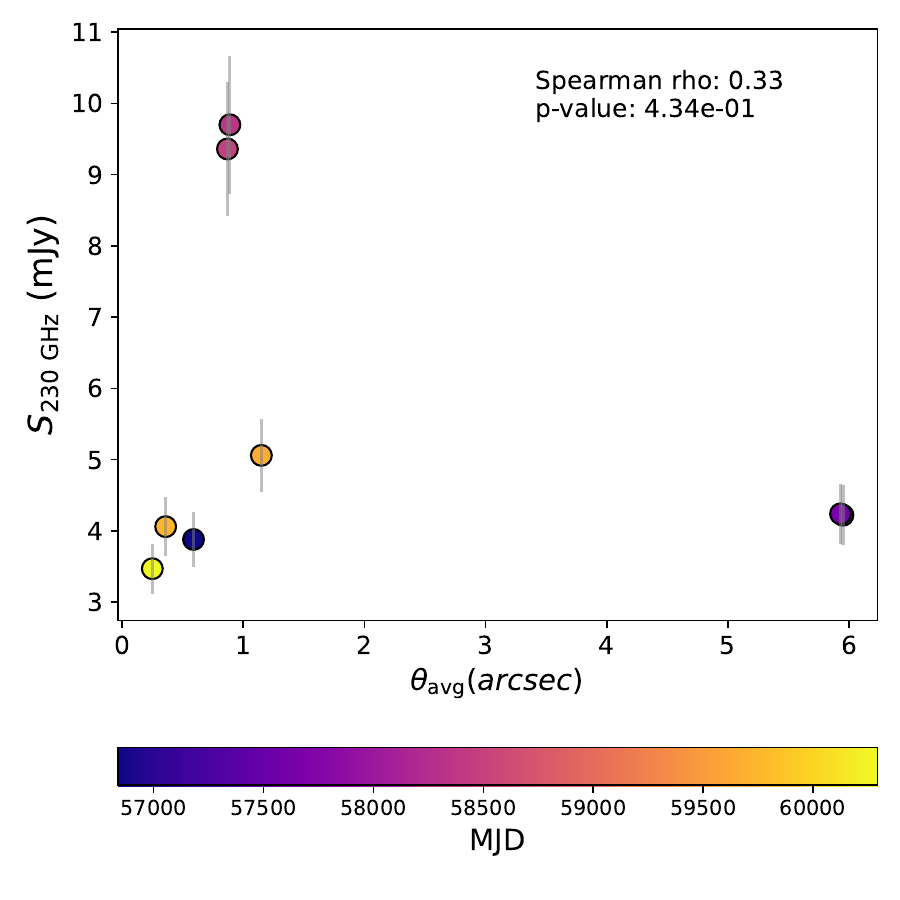}
\caption{Variation of the peak flux density ($S_{230~GHz}$) with average beam size ($\theta_{\rm avg}$).}
\label{fig:beam_flux}
\end{figure}

In order to maintain consistency of the spatial resolution across the different epochs, we made their final clean images with a same beam size of $7.31'' \times 4.96''$, which represents the lowest resolution observations.
We found the fluxes with the original beam size are within $\sim 5\%$ of the ones with the beam size of $7.3'' \times 4.96''$. Hence, we used the fluxes obtained from the original beam size in the paper.
Figure~\ref{fig:beam_flux} shows the variation of 230~GHz peak flux density ($S_{\rm 230~GHz}$) with the average beam size ($\theta_{\rm avg}$). The average beam size is calculated as $\theta_{\rm avg}=\sqrt{\theta_{\rm major}\times\theta_{\rm minor}}$. The observed flux is independent of the beam size, indicating our result is not affected by the different beam sizes.

\section{X-ray data analysis}
The $2-10$ keV fluxes were estimated by analyzing $0.5-10$~keV {\it Swift}/XRT spectra, processed using the UK Swift Science Data Centre tools \citep{Evans2009}. Spectra were modeled in \textsc{xspec} as \textsc{tbabs${\rm Gal}$ $\times$ zphabs $\times$ (BB+powerlaw)}, where \textsc{tbabs${\rm Gal}$} and \textsc{zphabs} account for Galactic and intrinsic absorption, and the \textsc{diskbb} and \textsc{powerlaw} components represent soft-excess and Comptonized emission, respectively.

% If you want to present additional material which would interrupt the flow of the main paper,
% it can be placed in an Appendix which appears after the list of references.

%%%%%%%%%%%%%%%%%%%%%%%%%%%%%%%%%%%%%%%%%%%%%%%%%%

%\bsp	% typesetting comment
%\label{lastpage}
\end{document}